\documentclass[twocolumn,showpacs,aps,prb,floatfix,superscriptaddress]{revtex4-1}

\usepackage{graphicx}

\begin{document}

\preprint{}

\title{Co-existing Singlet and Ordered $S=1/2$ Moments in the\\
 Ground State of the Triclinic Quantum Magnet CuMoO$_4$}

\author{S. Haravifard}\altaffiliation[Current affiliation:]{ The Advanced Photon Source, Argonne National Laboratory, Argonne, Illinois 60439, USA  and The James Franck Institute and Department of Physics, The University of Chicago, Chicago, Illinois 60637, USA.}
\affiliation{Department of Physics and Astronomy, McMaster University, Hamilton, Ontario, L8S 4M1, Canada}

\author{K. Fritsch}
\affiliation{Department of Physics and Astronomy, McMaster University, Hamilton, Ontario, L8S 4M1, Canada}

\author{T. Asano}
\affiliation{Department of Physics and Astronomy, McMaster University, Hamilton, Ontario, L8S 4M1, Canada}
\affiliation{Department of Physics, Kyushu University, Fukuoka 812-8581, Japan}

\author{J.P. Clancy}
\affiliation{Department of Physics and Astronomy, McMaster University, Hamilton, Ontario, L8S 4M1, Canada}

\author{Z. Yamani}
\affiliation{Canadian Neutron Beam Centre, NRC, Chalk River Laboratories, Chalk River, Ontario, K0J 1J0, Canada}

\author{G. Ehlers}
\affiliation{Neutron Scattering Science Division, Oak Ridge National Laboratory, Oak Ridge, Tennessee 37831-6475, USA}

\author{T. Nishimura}
\affiliation{Department of Physics, Kyushu University, Fukuoka 812-8581, Japan}

\author{Y. Inagaki}
\affiliation{Department of Applied Quantum Physics, Kyushu University, Fukuoka 812-0395, Japan}

\author{T. Kawae}
\affiliation{Department of Applied Quantum Physics, Kyushu University, Fukuoka 812-0395, Japan}

\author{I. Swainson}
\affiliation{Canadian Neutron Beam Centre, NRC, Chalk River Laboratories, Chalk River, Ontario, K0J 1J0, Canada}

\author{B.D. Gaulin}
\affiliation{Department of Physics and Astronomy, McMaster University, Hamilton, Ontario, L8S 4M1, Canada}
\affiliation{Brockhouse Institute for Materials Research, McMaster University, Hamilton, Ontario, L8S 4M1, Canada}
\affiliation{Canadian Institute for Advanced Research, 180 Dundas St. W., Toronto, Ontario, M5G 1Z8, Canada}

\begin{abstract}

CuMoO$_4$ is a triclinic quantum magnet based on $S=1/2$ moments at the Cu$^{2+}$ site. It has recently attracted interest due to the remarkable changes in its chromic and volumetric properties at high temperatures, and in its magnetic properties at low temperatures. This material exhibits a first order structural phase transition at $T_{\mathrm {C}}$ $\sim$ 190 K as well as a magnetic phase transition at $T_{\mathrm {N}}$ $\sim$ 1.75 K. We report low temperature heat capacity measurements as well as extensive elastic and inelastic neutron scattering measurements on powder samples taken above and below $T_{\mathrm {N}}$. We observe neutron diffraction consistent with a simple (1/2, 0, 0) antiferromagnetic structure indicating a doubling of the a-axis periodicity below $T_{\mathrm {N}}$. In addition, inelastic neutron scattering above a spin gap of $\sim$ 2.3 meV is consistent with triplet excitations out of paired $S=1/2$ moments which form singlet dimers. Low lying spin wave excitations are also observed and these originate from ordered $S=1/2$ moments below $T_{\mathrm {N}}$. Taken together these measurements show the ground state of CuMoO$_4$ to display both non-magnetic singlets, and ferromagnetically-coupled spins coexisting within an antiferromagnetic structure below $T_{\mathrm {N}}$ $\sim$ 1.75 K.

\end{abstract}


\maketitle

\section{Introduction}

Quantum magnets, based on $S=1/2$ local moments, have been of intense recent interest due to the nature of the exotic ground states they display and their relation to high temperature superconductivity\cite{1}. The ground states they can display are varied. One limiting case is that of a non-magnetic singlet state typical of quasi-one-dimensional spin-Peierls systems, MEM-(TCNQ)$_2$\cite{2}, CuGeO$_3$\cite{3,4}, TiOCl\cite{5,6,7} and TiOBr\cite{8,9}, as well as certain quasi-two dimensional Shastry-Sutherland systems, such as Sr$_2$Cu(BO$_3$)$_2$\cite{10}. However, antiferromagnetic N$\acute{e}$el ground states also exist, as occurs in the parent compounds of the high temperature superconductors, such as La$_2$CuO$_4$\cite{11,12,13}.

CuMoO$_4$ is a triclinic magnetic insulator made up of networks of quantum $S=1/2$ magnetic moments residing at the Cu$^{2+}$ site\cite{14,15}. Several different polymorphs of CuMoO$_4$ have been reported \cite{16,17,18,19,20}. At high pressure, CuMoO$_4$ crystallizes in two distorted wolframite-like structures, CuMoO$_4$-II and CuMoO$_4$-III, that both display antiferromagnetic order at low temperatures \cite{21,22}. Another polymorph, $\epsilon$-CuMoO$_4$, has a monoclinic crystal structure under ambient conditions and orders magnetically with a ferromagnetic component below $\sim$ 10 K \cite{23}.

The CuMoO$_4$ polymorphs which are the subject of the present article exhibit two triclinic phases at high and low temperatures and ambient pressure, the $\alpha$ and $\gamma$ phases, respectively. There is a strongly hysteretic 1st order structural phase transition between these two structures at $T_{\mathrm {C}}$ $\sim$ 190 - 250 K\cite{15,16,24}. While both the $\alpha$ (high temperature) and $\gamma$ (low temperature) phases are triclinic, they differ in unit cell volume by a remarkable 13\% on either side of $T_{\mathrm {C}}$, with the low temperature $\gamma$ phase displaying the smaller cell volume. This phase change is accompanied by a change in color of the material from green ($\alpha$) to red-ish brown ($\gamma$). For this reason, this material is referred to as displaying piezo or thermal chromism, and is of considerable current interest for these properties alone\cite{25,26,27}. 

The lattice parameters for CuMoO$_4$ in both its high temperature ($\alpha$) phase and its low temperature ($\gamma$) phase are listed in Table 1 as taken from \cite{15}. Along with the unit cell volume reduction of 13\% on cooling through $T_{\mathrm {C}}$, the lattice constant shrinks by $\sim$ 7\%, with the largest change being along the {\bf b}-axis\cite{15,16,24}.

\begin{table}[h]
\begin{tabular}{|c|c|}
\hline
$\alpha$-CuMoO$_4$&$\gamma$-CuMoO$_4$ \\
high temperature phase&low temperature phase\\\hline
space group: P$\bar{1}$ (No. 2)&space group: P$\bar{1}$ (No. 2)\\
a = 9.901(3) \AA&a = 9.699(9) \AA\\
b = 6.786(2) \AA&b = 6.299(6) \AA\\
c = 8.369(3) \AA&c = 7.966(7) \AA\\
$\alpha$ = 101.13(1)$^\circ$&$\alpha$ = 94.62(4)$^\circ$\\
$\beta$ = 96.88(1)$^\circ$&$\beta$ = 103.36(4)$^\circ$\\
$\gamma$ = 107.01(1)$^\circ$&$\gamma$ = 103.17(4)$^\circ$\\\hline
\end{tabular}
\caption{Lattice parameters for CuMoO$_4$ in its high temperature ($\alpha$) and its low temperature ($\gamma$) phases\cite{15}.}
\end{table}

The structure within the $\alpha$ phase can be described in terms of relatively isolated clusters of six Cu-O polyhedra, while within the $\gamma$ phase the Cu-O polyhedra take on a one-dimensional connectivity within the {\bf a-b} plane\cite{14}. The connectivity of the CuO$_6$ octahedra in its $\gamma$ phase gives rise to chains of ``molecules'' formed by 6 corner- and edge-sharing CuO$_6$ octahedra. Cu$^{2+}$ $S=1/2$ magnetic moments connected by corner-sharing octahedra are expected to interact via relatively strong antiferromagnetic exchange, due to the $\sim$ 180$^{\circ}$ Cu-O-Cu bond angles, while those which are connected by edge-sharing octahedra would experience relatively weaker magnetic coupling. The expectation arising from the connectivity within the six CuO$_6$ octahedra is that the six $S=1/2$ moments would form two loosely coupled singlet dimers and two relatively free $S=1/2$ moments per unit cell at low temperature\cite{14}. This scenario is also supported by magnetization measurements which show a magnetization plateau typical for singlet ground state systems \cite{28}. This paper reports on new heat capacity and neutron scattering measurements on polycrystalline CuMoO$_4$ which characterize its magnetic properties above and below $T_{\mathrm {N}}$ $\sim$ 1.75 K, within its $\gamma$ phase structure. As we will discuss, the scenario of moments arranged into two loosely coupled singlet dimers and two relatively free $S=1/2$ moments per unit cell is fully consistent with the measurements we present.

\section{Experimental Details}

Polycrystalline samples were prepared by mixing the following powders in stoichiometric proportions:

\begin{center}
MoO$_3$ + CuO $\rightarrow$ CuMoO$_4$	
\end{center}

\noindent
The mixed powders were pressed hydrostatically at 65~MPa, and annealed in air at 700$^\circ$C for 72 hours. Powder X-ray diffraction measurements of the final samples revealed high quality polycrystalline material with very little CuO residue.

In order to investigate the magnetic structure and magnetic excitations associated with the low temperature ground state of CuMoO$_4$, we carried out both elastic and inelastic neutron scattering measurements on polycrystalline samples as a function of temperature and magnetic field on the C2 powder diffractometer and the C5 triple axis spectrometer at the Canadian Neutron Beam Centre (CNBC), Chalk River, as well as inelastic time-of-flight neutron scattering measurements using the Cold Neutron Chopper Spectrometer (CNCS)\cite{21} at the Spallation Neutron Source (SNS) at Oak Ridge National Laboratory (ORNL).

The polycrystalline samples used for the measurements at CNBC were loaded in a sealed Al can in a $^4$He exchange gas and mounted in different cryostats: either a pumped $^3$He or $^4$He cryostat, in order to achieve temperatures as low as 0.3 K and 1.5 K, respectively, and with magnetic field capabilities up to 7.5 T. For the triple axis measurements at CNBC, we employed a vertically focusing pyrolytic graphite (PG) monochromator and a flat analyzer with a fixed final energy of E$_f$ = 14.7 meV. Two PG filters were used in the scattered neutron beam in order to eliminate higher order wavelength contamination. A liquid N$_2$ cooled sapphire filter was used in the main beam to minimize the fast neutron background. Soller slits in the four beam paths (from source to detector) produced a collimation of [none, 0.48$^\circ$, 0.56$^\circ$, 1.2$^\circ$] resulting in an energy resolution of $\sim$ 1 meV for these triple axis measurements.

Time-of-flight neutron scattering measurements were performed using the CNCS at the SNS at ORNL. CNCS is a direct geometry, multi-chopper inelastic spectrometer optimized for high-{\bf Q} and good-E resolution measurements using low incident neutron energies (E$_i$ $<$ 30 meV). Measurements were carried out on a 17 g polycrystalline CuMoO$_4$ sample that was loaded in a standard aluminum sample can. The sample environment consisted of a 5 T vertical field magnet cryostat with a base temperature of 1.5 K. A cadmium mask was attached to the sample can to reduce multiple scattering and scattering from the cryostat. The sample was measured using two different settings for the incident neutron energy to cover a broad range of (Q, E) space.
E$_i$ = 6.6 meV and E$_i$ = 1.55 meV were employed for moderate and high energy resolution measurements, respectively. For all measurements, the spectrometer was run in the high-flux configuration and the double-disk chopper was phased at 180 Hz, providing elastic energy resolutions of $\sim$ 0.18 meV, and $\sim$~0.025~meV for the two E$_i$ settings, respectively.

\section{Heat Capacity Measurements}

\begin{figure}[b]
\includegraphics{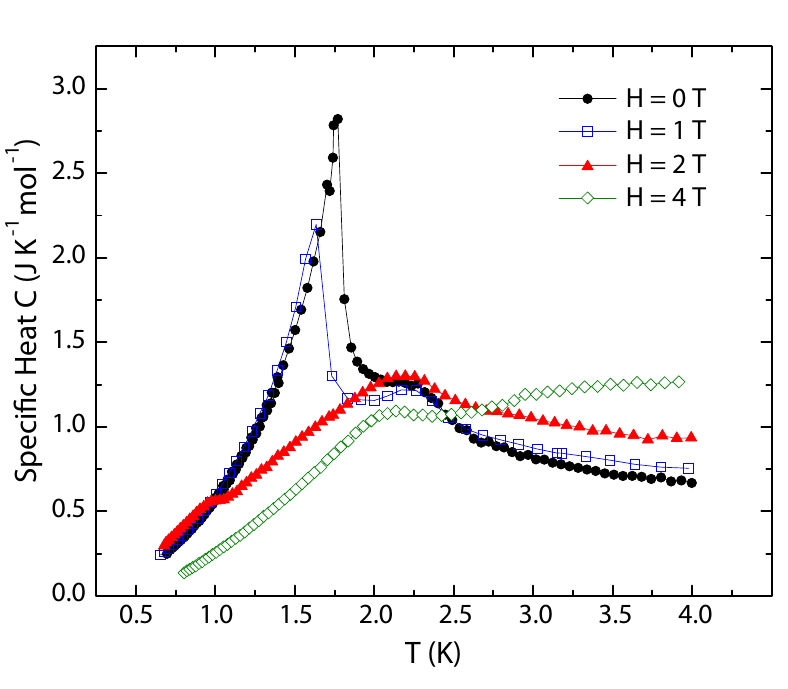}
\caption{\label{1}(Color online) The temperature dependence of the heat capacity of CuMoO$_4$ as observed in zero magnetic field as well as in magnetic fields up to 4 T.}
\end{figure}

We carried out heat capacity measurements at low temperatures to characterize the magnetic phase behavior in CuMoO$_4$. Figure 1 shows heat capacity measurements as a function of temperature and magnetic field, up to 4 T. The measurements were performed using the quasi-adiabatic heat pulse method and a $^3$He refrigerator. The heat capacity measurements at H=0 T reveal a $\lambda$-like anomaly signifying a magnetic phase transition at $T_{\mathrm {N}}$ $\sim$ 1.75 K, as well as a weaker peak near $\sim$ 2.2 K, likely indicating a buildup of short-range correlations at that temperature. The sharp $\lambda$-like anomaly moves to lower temperatures and weakens in amplitude on application of a magnetic field. The transition appears to have been fully suppressed for fields $>$ 2 T, with only a broad anomaly remaining in the heat capacity at $\sim$ 2.2 K for higher fields.

\section{Neutron Scattering Results and Discussion}

We performed two sets of neutron powder diffraction measurements on CuMoO$_4$. The first of these, shown in Fig. 2 (a), was performed on the C2 powder diffractometer at CNBC, Chalk River, using an incident neutron wavelength of $\lambda$ = 2.37 \AA. Figure 2 (a) shows the low angle portion of the neutron diffraction pattern at T = 5 K and T = 0.4 K, and a clear temperature dependent, resolution-limited Bragg peak is observed at a scattering angle of $\sim$ 7.4$^{\circ}$ for which $Q$ = 0.342 \AA$^{-1}$. This was the only additional Bragg peak to appear on cooling through $T_{\mathrm {N}}$ $\sim$ 1.75 K.

\begin{figure}[h]
\includegraphics{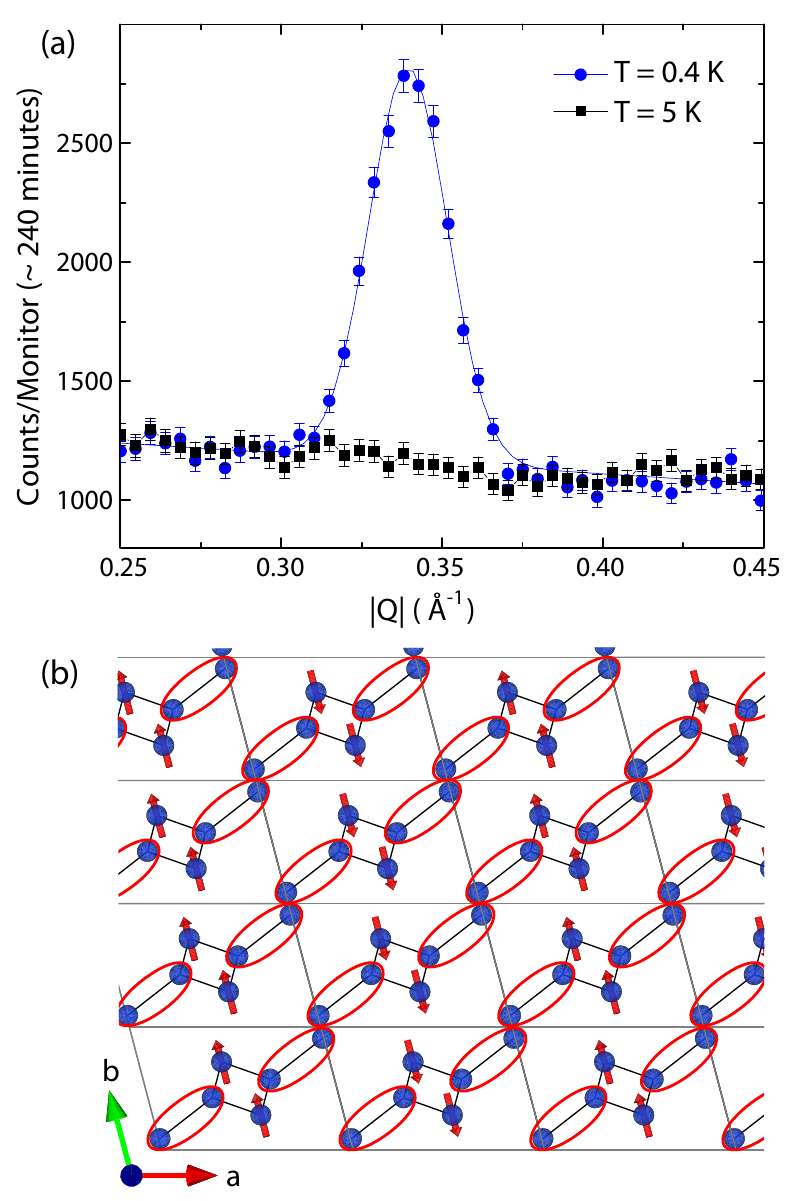}
\caption{\label{3}(Color online) (a) Elastic neutron scattering measured at the C2 powder diffractometer at CNBC showing the magnetic Bragg peak at $|$Q$|$ $\sim$ 0.34 \AA$^{-1}$ at $\sim$ 0.4 K. A complete powder diffraction pattern using the position-sensitive linear detector took $\sim$ 240 minutes. (b) The proposed spin configuration in the H = 0 T ground state of CuMoO$_4$, based on the ordering wavevector of antiferromagetic arrangements of spins in neighboring cells consistent with the (1/2, 0, 0) ordering wavevector. Inelastic scattering indicates the presence of triplet excitations out of paired singlets, shown as ellipses surrounding the $S=1/2$ moments which pair to form the singlets and which take part in the magnetic structure.}
\end{figure}

Powder diffraction measurements were also performed on the C5 triple axis spectrometer which allowed a parametric study of the temperature and magnetic field dependence of the low temperature Bragg peak at Q = 0.342 \AA$^{-1}$ shown in Fig. 2 (a). These order parameter measurements are shown in Fig. 3 (a) and (b), for the temperature dependence at zero field, and the field dependence at T = 0.4 K, respectively. This field and temperature dependence of the order parameter identify the new low temperature Bragg peak as magnetic in origin and corresponding to the sharp anomaly observed in the zero field heat capacity, as shown in Fig. 1. Interestingly, the phase transition appears to be continuous as a function of temperature at zero field (Fig. 3 (a)) but rather discontinuous as a function of field at low temperatures (Fig. 3 (b)).

While we have observed only a single magnetic Bragg peak for CuMoO$_4$ below $T_{\mathrm {N}}$, we can model its magnetic structure, based on its periodicity as shown in Fig. 2 (b), wherein we pair off 4 of the 6 $S=1/2$ moments per unit cell into 2 singlets, and allow the remaining 2 spins to order ferromagnetically within a unit cell and antiferromagnetically from cell to cell along the triclinic {\bf a} direction. The rationale for pairing 4 of the 6 $S=1/2$ spins per unit cell off into non-magnetic singlets comes both from recent high-field magnetization studies \cite{28} and from inelastic neutron scattering results which we report below. The magnitude of the ordering wavevector Q = 0.34 \AA$^{-1}$ is correctly accounted for by the magnitude of the resulting (1/2, 0, 0) antiferromagnetic ordering wavevector within this triclinic structure.

\begin{figure}
\includegraphics{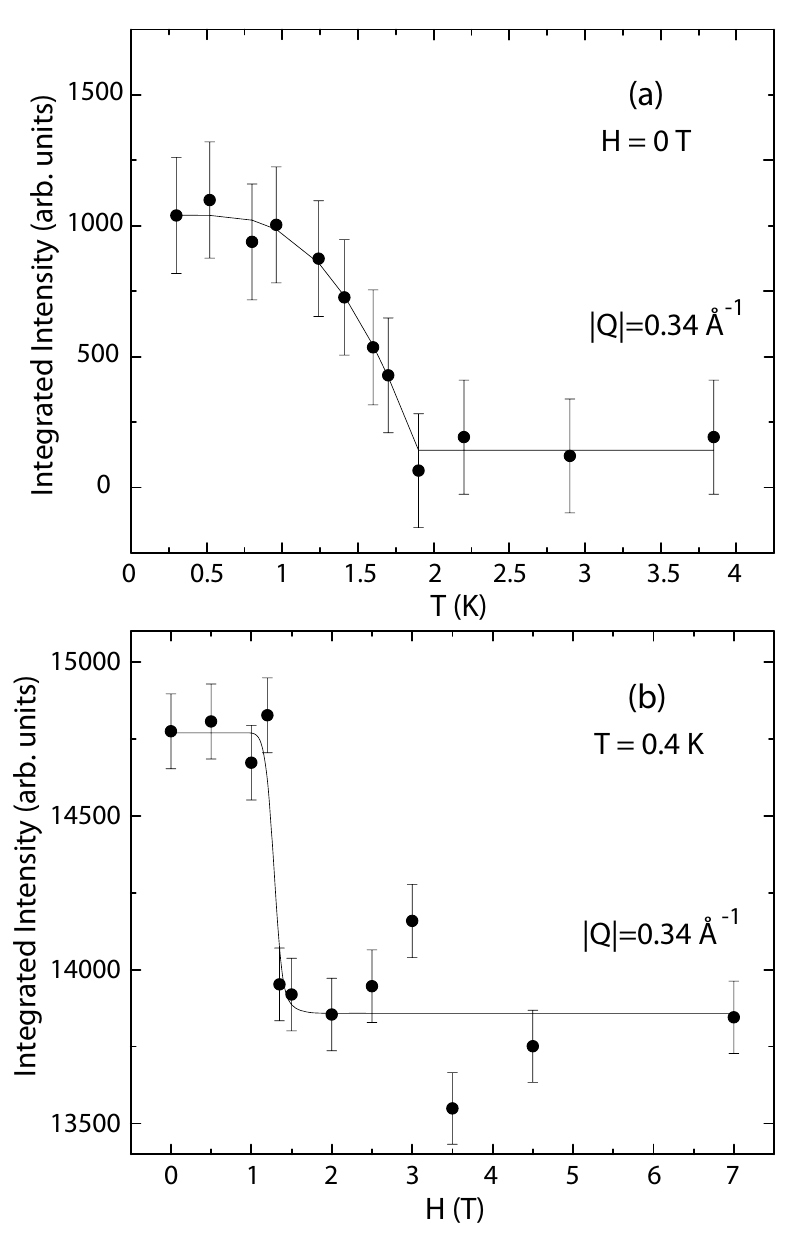}
\caption{\label{4} (a) The antiferromagnetic order parameter as measured by the integrated intensity of the Q = 0.34 \AA$^{-1}$ Bragg peak is shown as a function of temperature. A high temperature (T = 50 K) background has been subtracted from this data set. (b) The same antiferromagnetic Bragg intensity at Q = 0.34 \AA$^{-1}$ is shown as a function of magnetic field at T $\sim$ 0.4 K, within the magnetically ordered state. While the thermal evolution of the transition appears continuous in zero field, the antiferromagnetic order drops discontinuously to zero with field at low temperatures. The lines in both (a) and (b) are guides to the eye, with error bars of $\sigma$ from counting statistics. Both data sets shown are obtained on the triple axis spectrometer C5 at the CNBC.}
\end{figure}

We also carried out two sets of inelastic neutron scattering measurements on this polycrystalline sample. We will first describe time-of-flight inelastic measurements taken with the CNCS chopper spectrometer at SNS, and then triple axis measurements taken with the C5 spectrometer at CNBC, Chalk River.

The inelastic excitation spectrum for CuMoO$_4$ is shown in Fig. 4 as a color contour map of the inelastic neutron scattering intensity, S(Q, E), for an incident neutron energy of E$_i$ = 6.6 meV. Figure 4 (a), (b), and (c) show this spectum in the ordered phase below $T_{\mathrm {N}}$ at T = 1.5 K and at zero applied magnetic field; in the disordered phase at base temperature T = 1.5 K and H = 4 T; and in the paramagnetic phase at H = 0 T and T = 6 K, respectively. Figure 4 (d) shows a cut in energy, of these same three data sets, integrated in Q for $|Q| = [0.6, 1.0]$ \AA$^{-1}$. These data sets have had a high temperature, 50 K, H = 0 T background data set subtracted from them.

These inelastic data sets are consistent with low energy spin wave excitations in the ordered state below $T_{\mathrm {N}}$ (Fig. 4 (a)) coexisting with a gapped excitation spectrum characteristic of triplet excitations out of a singlet ground state. The triplet excitations have a band width of $\sim$ 2.5 meV, with a gap of $\sim$ 2.3 meV. The band width likely originates from weak dispersion within the triplet of excited states due to inter-dimer exchange coupling, as observed in other singlet ground state systems such as SrCu$_2$(BO$_3$)$_2$\cite{10}.

On applying an H = 4 T magnetic field at low temperatures (Fig. 4 (b)), the bottom of the triplet band moves down to below $\sim$ 1.9 meV, consistent with an expected downward shift of the energy of the lowest of the three triplet states by $\Delta E = g\mu_BH=$0.46 meV for g=2 and H=4 T. The low energy spin wave scattering is also raised in energy, and the spin wave spectrum appears to be gapped. This can be seen more clearly in the energy cuts of this same data shown in Fig. 4 (d), wherein the low energy spectrum below $\sim$ 0.5 meV for T = 1.5 K and H = 0 T, appears to be depleted and it displays a pronounced inelastic peak at $\sim$ 0.8 meV for T = 1.5 K and H = 4 T.

\begin{figure}[h]
		\includegraphics{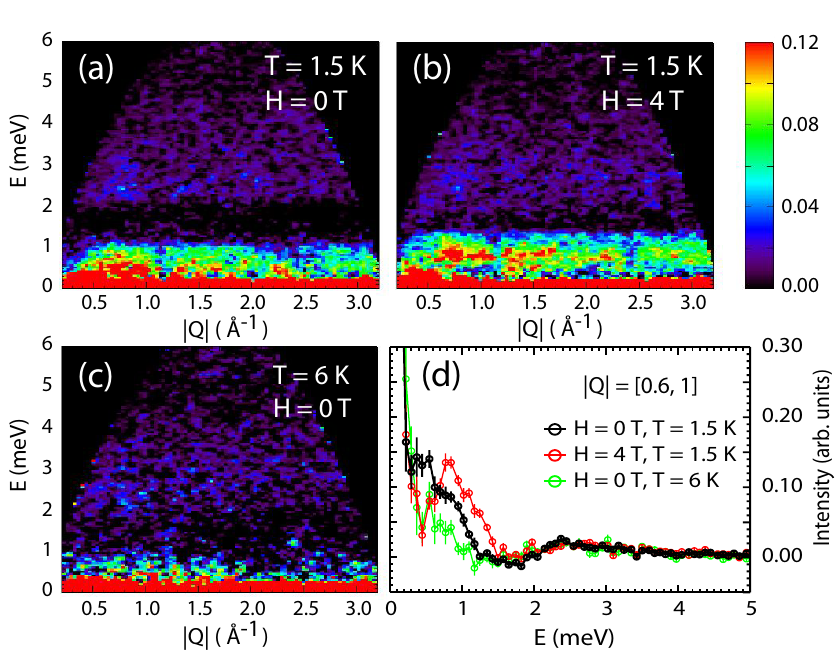}
\caption{(Color online) Color contour maps of S(Q, E) observed in CuMoO$_4$ are shown for E$_i$ = 6.6 meV at T = 1.5 K in zero applied magnetic field (a) and in an H = 4 T magnetic field (b). Panel (c) shows S(Q, E) in the H = 0 T paramagnetic phase at T = 6 K. An energy cut through S(Q, E) for an interval of $|Q|$ = 0.6 to 1 \AA$^{-1}$. A high temperature background (at T = 50 K, within the paramagnetic state) has been subtracted from all panels to isolate the magnetic scattering.}
\end{figure}

Higher energy resolution measurements were also taken with CNCS using an incident neutron energy of E$_i$ = 1.55~meV, and these are shown for T = 1.5 K in Fig. 5. Again a high temperature data set at T = 50 K and H = 0 T has been subtracted from this data set to isolate the magnetic scattering from the sample. This higher energy resolution data set clearly shows a Goldstone spin wave mode emanating out of the magnetic Bragg peak position at Q = 0.34 \AA$^{-1}$. Beyond $\sim$ 0.6 \AA$^{-1}$ the spin wave density of states is strongly peaked at $\sim$ 0.55 meV, however, a second distribution of magnetic scattering, with a bandwidth of $\sim$ 0.25 meV is evident between 0.7 and 1.0 meV. These two bands of spin wave scattering account for the quasi-elastic magnetic scattering observed with lower energy resolution (Fig. 4) wherein quasi-elastic scattering is observed out to $\sim$ 1 meV.  Given that the magnetically ordered ground state below $T_{\mathrm {N}}$ is described by an antiferromagnetic alternation of ferromagnetically-coupled spins along the trigonal {\bf a} direction, it is not surprising that two bands of spin wave excitations, an acoustic and an optic band, would be present at low temperatures. 

\begin{figure}[h]
		\includegraphics{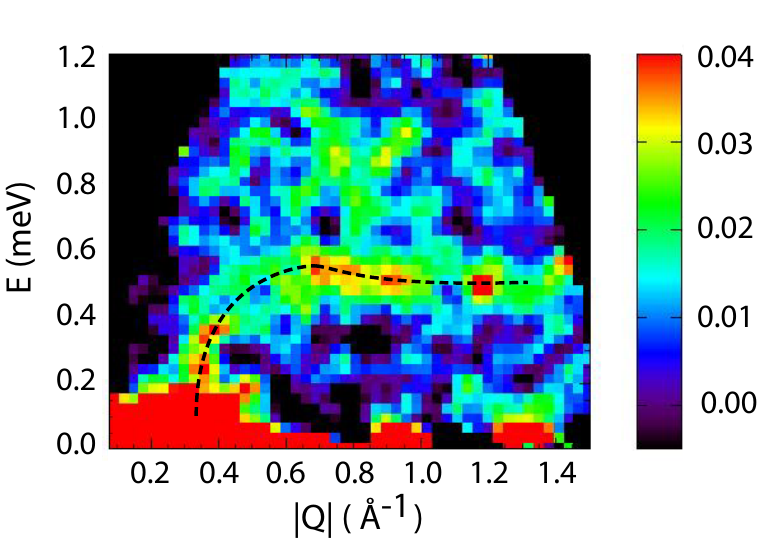}
	\caption{(Color online) A high energy resolution measurement of S(Q, E) using CNCS and an incident neutron energy of E$_i$ = 1.55 meV is shown for CuMoO$_4$ at T = 1.5 K and H = 0 T. A high temperature (T = 50 K) background data set has been subtracted from the data set. The low energy spin dynamics are seen to consist of a Goldstone mode emanating from the ordering wavevector, Q = 0.34 \AA$^{-1}$, and a relatively dispersionless band of excitations near 0.5 meV. Magnetic spectral weight is observed out to $\sim$ 1 meV.}
\end{figure}

\begin{figure}[h]
\includegraphics{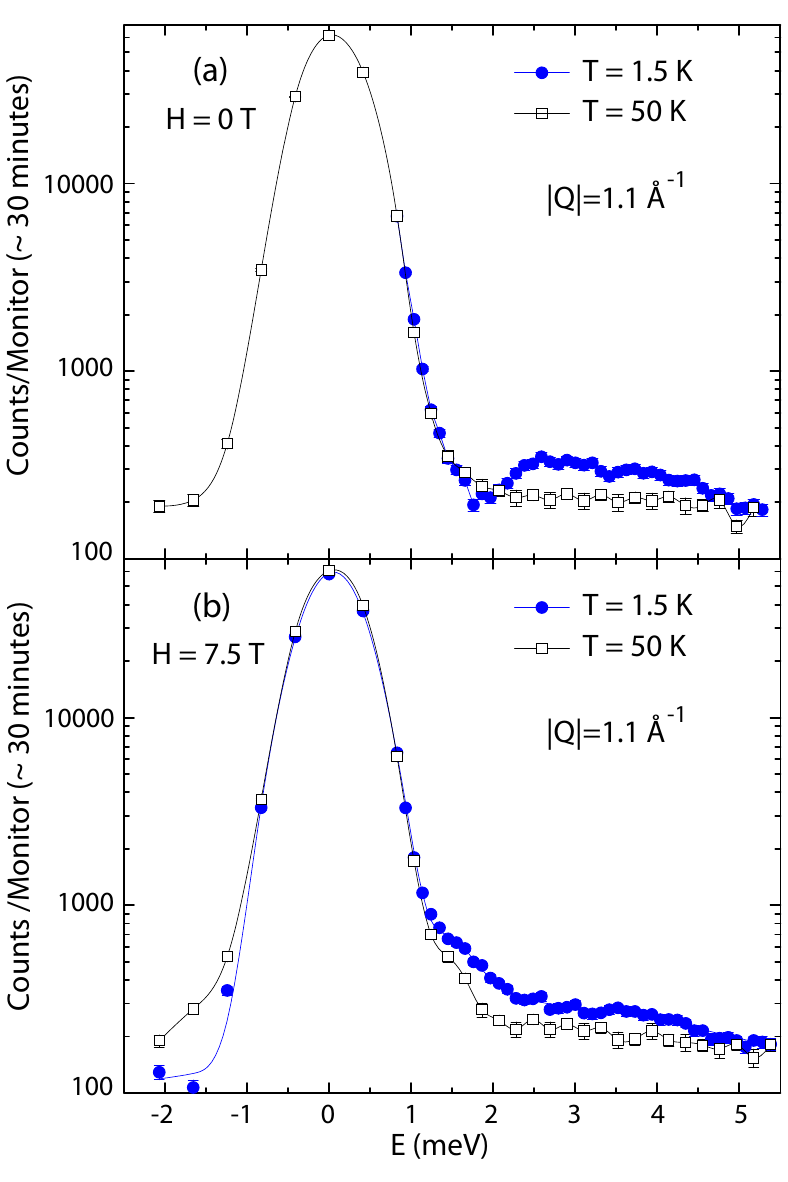}
\caption{\label{6}(Color online) (a) Relatively low energy resolution inelastic scattering at Q = 1.1 \AA$^{-1}$ reveals a gapped spin excitation spectrum in the ordered state at T = 1.5 K and in zero magnetic field. The gap is $\sim$ 2.3 meV with a bandwidth of $\sim$ 2.5 meV. (b) At the largest applied magnetic field, we observe the lower bound of the zero field triplet band to extend well within the $\sim$ 2.3 meV gap. Note the logarithmic intensity scale.}
\end{figure}

\begin{figure}[h]
\includegraphics{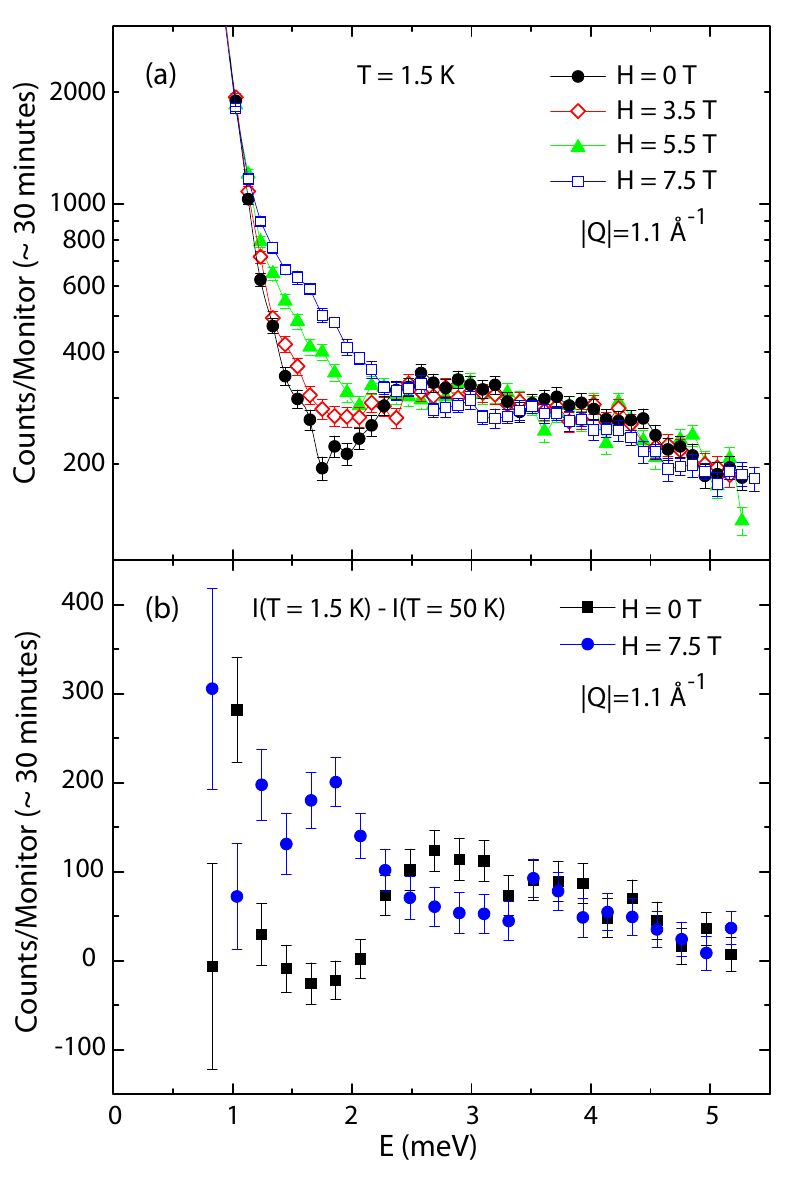}
\caption{\label{7}(Color online) (a) The effect of application of a magnetic field in the range of 0 T to 7.5 T on the magnetic excitation spectrum is shown in CuMoO$_4$. (b) The high temperature background-subtracted inelastic intensity at zero field and at an applied field of 7.5 T is shown. Clearly the bottom of the triplet scattering near $\sim$ 2.5 meV is depleted and transferred down to $\sim$ 1.75 meV, within the gap, by application of the magnetic field. This is consistent with expectations for Zeeman splitting of a weakly dispersive triplet band. Simultaneously, the spectral weight of the low energy spin wave excitations are moved to higher energies within the gap. Note that the intensity scale is logarithmic in (a) and linear in (b).}
\end{figure}

Constant-Q = 1.1 \AA$^{-1}$ inelastic neutron scattering measurements carried out with the C5 triple axis spectrometer at CNBC, Chalk River are shown in Figs. 6 and 7. These largely corroborate the time-of-flight measurements show in Figs. 4 and 5, although they are carried out to larger applied magnetic fields, up to H = 7.5~T. These measurements were performed with an energy resolution of $\sim$ 1 meV, and thus detail within the quasi-elastic spin wave band of scattering is obscured by nuclear elastic incoherent scattering. Nonetheless, the triplet excitations and singlet-triplet gap can be clearly observed, as well as the filling of the gap with increasing magnetic field.

Figure 6 (a) shows S(Q = 1.1 \AA$^{-1}$, E) for T = 1.5 K and T = 50 K  and zero magnetic field, while Fig. 6 (b) shows the same spectra but for H = 7.5 T. Figure 7 (a) shows the field dependence of S(Q = 1.1 \AA$^{-1}$, E) at T = 1.5 K, while Fig. 7 (b) focusses on the low energy part of this spectrum, for H = 0 T and H = 7.5 T only, after the high temperature background at T = 50 K has been subtracted from both data sets. Note that the intensity scale for Fig. 6 (a) and (b) and Fig. 7 (a) is logarithmic, while that of Fig. 7 (b) is linear. This comparison clearly shows spectral weight from the low energy side of the triplet band being depleted and displaced downwards in energy by g$\mu_B$H $\sim$ 0.7 meV to $\sim$ 1.6 meV. The low energy spin waves are also raised in energy by application of a field, and this also pushes magnetic intensity into the gap. The higher end of the triplet bandwidth, above 3.5 meV, shows less field dependence, but the scattering is weaker here, and the averaging effects of the triplet dispersion may be greater.

\section{Conclusions}

We have carried out heat capacity and both elastic and inelastic neutron scattering measurements on powder samples of the triclinic quantum magnet CuMoO$_4$ with the purpose of understanding the nature of its low temperature ground state. All results are consistent with an antiferromagnetic long range ordered ground state appearing below $T_{\mathrm {N}}$ = 1.75 K in H = 0 T. The ordering wavevector associated with this antiferromagnetic order is identified as (1/2, 0, 0), and this is consistent with a low temperature state in which the molecule of six Cu$^{2+}$ ions, which makes up the triclinic structure, organizes into 4 $S=1/2$ moments which form two singlets, as well as 2 ferromagnetically coupled $S=1/2$ moments which then order antiferromagnetically along the triclinic {\bf a} direction. We explicitly show that at low temperatures this ordered state is destroyed by an applied magnetic field of $\sim$ 1.5 T.

Our inelastic neutron scattering measurements on CuMoO$_4$ probe the magnetic excitation spectrum and show it to be well described by dispersive triplet excitations with a gap of $\sim$ 2.3 meV and a bandwidth of $\sim$ 2.5 meV. Low lying spin wave excitations are also observed, and are shown to display a Goldstone mode for T$<$ $T_{\mathrm {N}}$ which is soft at the ordering wavevector of Q $\sim$ 0.35 \AA$^{-1}$, as well as a second branch of spin wave excitations forming a band in the approximate range 0.6 - 1.0 meV.

The picture arising from the measurements we present, of an antiferromagnetic long range ordered structure made up of co-existing non-magnetic singlets and ferromagnetically-coupled spins, is clearly exotic, comprised as it is by both of the ground states normally associated with antiferromagnetism in materials. CuMoO$_4$ is clearly an interesting and exotic example where both ordered spins and singlets are the building blocks from which the antiferromagnetic ground state is constructed. We hope this study will guide and inform further theoertical studies of this and related quantum magnets.

\section{Acknowledgements}
We wish to acknowledge the contributions of K.A. Ross and J.P.C. Ruff to the neutron scattering measurements reported here. This work was supported by NSERC of Canada and by the Research Exchange Program between JSPS and NSERC and Grants-in-Aid for Scientific Research, MEXT (No.21560723).

\end{document}